\documentclass[showpacs  ,secnumarabic, nobibnotes,  nofootinbib, aps,
prd]{revtex4}%
\usepackage{amsmath}
\usepackage{amssymb}
\usepackage{bm}
\usepackage{graphicx}
\usepackage{amsfonts}%
\providecommand{\U}[1]{\protect\rule{.1in}{.1in}}

\begin{document}
\title{Role of gluons and the quark sea in the proton spin 
\footnote{Accepted to Phys. Lett. B}}
\author{Petr Zavada}
\email{zavada@fzu.cz}
\affiliation{Institute of Physics AS CR, Na Slovance 2, CZ-182 21 Prague 8, Czech Republic}

\begin{abstract}
The real, interacting elementary particle always consits of a 'bare' particle
and a cloud of virtual particles mediating a self-interaction and/or the bond
inside a composite object. In this letter we discuss the question of spin
content of the virtual cloud in two different cases: electron and quark.
Further, the quark spin is discussed in the context of proton spin, which is
generated by the interplay of quarks and virtual gluons. We present a general
constraint on the gluon contribution and make a comparison with the
experimental data.

\end{abstract}

\maketitle

\section{Introduction}

In our recent paper \cite{Zavada:2013ola} we studied the proton spin structure
in leading order of the covariant approach assuming the gluon contribution to
the proton spin can be neglected. However the question of the real role of
gluons in generating the proton spin is still open. Actually some recent data
obtained at RHIC and their analyses
\cite{Adare:2014hsq,Adamczyk:2014ozi,deFlorian:2014yva,Nocera:2014gqa,Bourrely:2014uha}
can suggest a positive gluon contribution to the proton spin. The main aim of
the present letter is to extend discussion from our previous study to the case
of nonzero gluon contribution. We will show what constraint on the gluon
contribution follows from the covariant approach.

Sec. \ref{spin} is devoted to a discussion about some general aspects of
particle spin and its scale dependence. Two different examples are considered,
electron and quark. The electron spin structure is also an interesting topic,
see the recent study \cite{Liu:2014fxa} and the previous papers
\cite{Brodsky:2000ii,Burkardt}. Particularly important questions concern the
spin of quarks inside the nucleon. In Sec. \ref{proton} the discussion about
the proton spin, which is generated by the interplay of angular moments of
quarks and gluons, continues in the context of recent experimental data.

In present calculations we use, as before, the rest frame of the composite
system as a starting reference frame. This frame is suitable for the
consistent composition of spins and OAMs of the constituents in the
representation of spinor spherical harmonics. The resulting state serves as an
input for construction of the covariant quantities, like the spin vector or
the spin structure functions \cite{Zavada:2013ola}.

\section{Spin of the particle in its scale dependent picture}

\label{spin}In general, description of real interacting particles can be
related to their 'bare' or 'dressed' form. In our present discussion we
address the general questions:

a) How much do the virtual particles surrounding bare particle contribute to
the spin of corresponding real, dressed particle?

b) How much do the virtual particles mediating bond of the constituents of a
composite particle contribute to its spin?

In quantum mechanic the total angular momentum (AM) of any particle including
composite ones is given by the sum of the orbital AM (OAM) and spin,
$\mathbf{J=L+S}.$ The corresponding quantum numbers are discrete sets of
integers or half-integers and in the relativistic case only total AM
conserves, so only $J$ and $J_{z}$ can be the good quantum numbers. We will
illustrate the problem with two different examples, electron and quark.

\subsection{Spin of electron}

\begin{figure}[t]
\includegraphics[width=6cm]{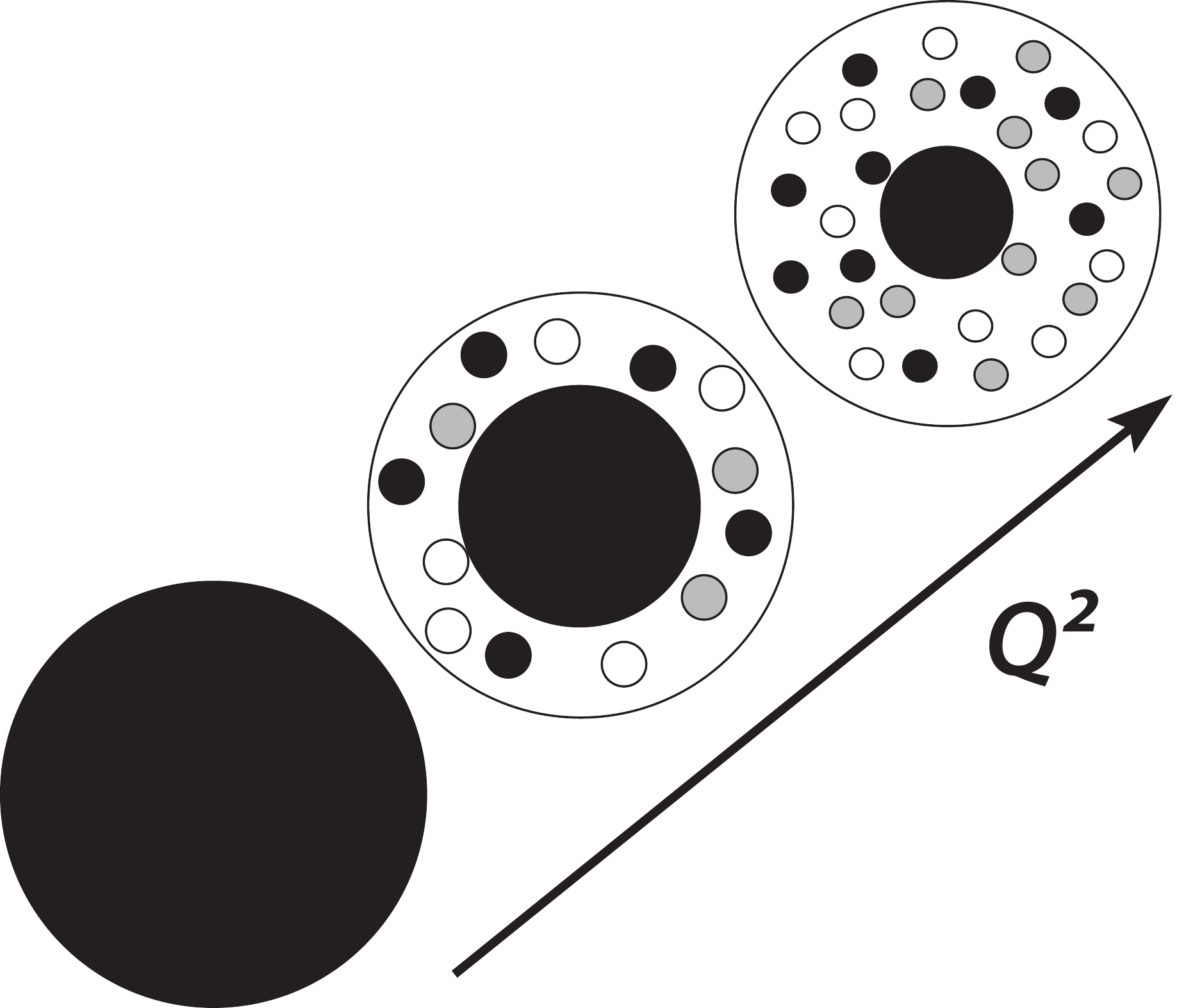}\caption{Scale dependent image of a
real particle, see text.}%
\label{fgr1}%
\end{figure}

The electron, as a Dirac particle, in its rest frame has AM defined by its
spin, $s=1/2$. This value is the same for the dressed electron (as proved
experimentally) and for the bare one (as defined by the QED Lagrangian). The
dressed electron is a bare electron surrounded by the virtual cloud of
$\gamma$ and $e^{-}e^{+}$ pairs, as symbolically sketched in Fig. \ref{fgr1}
for different scales represented by the parameter $Q^{2}$. So the
renormalization as a continuous change of the scale should not change the AM
represented by the discrete numbers, $J^{e}(Q^{2})=s=1/2$. But what about the
projections $J_{z}^{e}(Q^{2})+$ $J_{z}^{\gamma}(Q^{2})=\pm1/2$? Can the
contribution of virtual cloud $J_{z}^{\gamma}(Q^{2})$ differ from zero and how
much? \ In this letter we present a semiclassical estimate of the vector
$\mathbf{J}^{\gamma}$. The electromagnetic field, or its $\gamma$-quanta, are according
to Maxwell equations created by the electric current. We consider the current
generated by the electron states represented by the spinor spherical harmonics%
\begin{equation}
\left\vert j,j_{z}\right\rangle =\Phi_{jl_{p}j_{z}}\left(  \mathbf{r}\right)
=\frac{1}{\sqrt{2\epsilon}}\left(
\begin{array}
[c]{c}%
\sqrt{\epsilon+m}R_{kl_{p}}\Omega_{jl_{p}j_{z}}\left(  \mathbf{\omega}\right)
\\
-\sqrt{\epsilon-m}R_{k\lambda_{p}}\Omega_{j\lambda_{p}j_{z}}\left(
\mathbf{\omega}\right)
\end{array}
\right)  , \label{rs1}%
\end{equation}
where $\mathbf{\omega}$ represents the polar and azimuthal angles
($\theta,\varphi$) of the space coordinates $\mathbf{r}$ with respect\ to the
axis of quantization $z,$ $l_{p}=j\pm1/2,\ \lambda_{p}=2j-l_{p}$ ($l_{p}$
defines the parity), energy $\epsilon=\sqrt{\mathbf{k}^{2}+m^{2}}$ and
\begin{align}
\Omega_{jl_{p}j_{z}}\left(  \mathbf{\omega}\right)   &  =\left(
\begin{array}
[c]{c}%
\sqrt{\frac{j+j_{z}}{2j}}Y_{l_{p},j_{z}-1/2}\left(  \mathbf{\omega}\right) \\
\sqrt{\frac{j-j_{z}}{2j}}Y_{l_{p},j_{z}+1/2}\left(  \mathbf{\omega}\right)
\end{array}
\right)  ,\label{rs1e}\\
\Omega_{j\lambda_{p}j_{z}}\left(  \mathbf{\omega}\right)   &  =\left(
\begin{array}
[c]{c}%
-\sqrt{\frac{j-j_{z}+1}{2j+2}}Y_{\lambda_{p},j_{z}-1/2}\left(  \mathbf{\omega
}\right) \\
\sqrt{\frac{j+j_{z}+1}{2j+2}}Y_{\lambda_{p},j_{z}+1/2}\left(  \mathbf{\omega
}\right)
\end{array}
\right)  ,\nonumber
\end{align}
where $l_{p}=j-1/2$ and $\lambda_{p}=j+1/2$. The functions $Y_{l,l_{z}}$\ are
usual spherical harmonics. The radial functions in the case of free electron
read:%
\begin{align}
R_{kl}(r)  &  =\sqrt{\frac{2\pi k}{r}}J_{l+1/2}(kr),\label{rs2}\\
\int r^{2}R_{kl}R_{k^{\prime}l}dr  &  =2\pi\delta(k-k^{\prime}),\nonumber
\end{align}
where $k=\left\vert \mathbf{k}\right\vert $ and $J_{\nu}(z)$\ are Bessel
functions of the first kind, otherwise (e.g. for electron in the hydrogen
atom) the radial functions differ according to an external field. However, it
is important that the information about the electron AM is completely absorbed
in the angular terms and does not depend on the radial functions. The states
$\left\vert j,j_{z}\right\rangle $ are eigenstates of the total AM and have
been discussed before \cite{Zavada:2013ola} in momentum representation, while
in the present note we deal with their coordinate representation corresponding
to the rest frame of an composite system. The corresponding current reads%
\begin{equation}
I_{\mu}=(I_{0},\mathbf{I})=\Phi_{jl_{p}j_{z}}^{\dagger}\left(  \mathbf{r}%
\right)  \gamma^{0}\gamma_{\mu}\Phi_{jl_{p}j_{z}}\left(  \mathbf{r}\right)
\label{br3}%
\end{equation}
and one can check that%
\begin{equation}
I_{0}=h_{I}\rho_{j,j_{z}}\left(  \cos\theta\right)  ,\qquad\mathbf{I}%
=h_{II}\rho_{j,j_{z}}\left(  \cos\theta\right)  \mathbf{r}, \label{br4}%
\end{equation}
where%
\begin{align}
h_{I}  &  =\frac{1}{2}\left(  \left(  1+\frac{m}{\epsilon}\right)  R_{kl_{p}%
}^{2}+\left(  1-\frac{m}{\epsilon}\right)  R_{k\lambda_{p}}^{2}\right)
,\label{br6}\\
h_{II}  &  =-\frac{k}{\epsilon r}R_{kl_{p}}R_{k\lambda_{p}}.\nonumber
\end{align}
A few examples of $\rho_{j,j_{z}}$\ are given in Table \ref{tb1}, where one
can observe the following. \begin{table}[pt]
\begin{center}%
\begin{tabular}
[c]{c|c|}%
$j,j_{z}$ & $\rho_{j,j_{z}}(\omega)$\\\hline
$\frac{1}{2},\pm\frac{1}{2}$ & $1$\\
$\frac{3}{2},\pm\frac{3}{2}$ & $\frac{3-3\cos2\theta}{4}$\\
$\frac{3}{2},\pm\frac{1}{2}$ & $\frac{5+3\cos2\theta}{4}$\\
$\frac{5}{2},\pm\frac{5}{2}$ & $\frac{45-60\cos2\theta+15\cos4\theta}{64}$\\
$\frac{5}{2},\pm\frac{3}{2}$ & $\frac{57-12\cos2\theta-45\cos4\theta}{64}$\\
$\frac{5}{2},\pm\frac{1}{2}$ & $\frac{45+36\cos2\theta+15\cos4\theta}{32}$%
\end{tabular}
\end{center}
\caption{The examples of the angular distributions $\rho_{j,j_{z}}$. The
common factor $1/4\pi$ is omitted. }%
\label{tb1}%
\end{table}The stationary current $I_{\mu}$ depends only on $j$ and
$\left\vert j_{z}\right\vert $, therefore it does not involve any information
on the direction of electron polarization. So, there is no reason to expect
any correlation between electron polarization and polarization of the
electromagnetic field generated by this current, or equivalently polarization
of the statistical set of emitted and reabsorbed $\gamma$. In other words the
average polarization of virtual cloud of $\gamma$ and consequently also
$e^{-}e^{+}$ pairs should be zero. The AM of the electromagnetic field is
given by the relation%
\begin{equation}
\mathbf{J}^{\gamma}=\int\mathbf{r}\times(\mathbf{E}\times\mathbf{H}%
)d^{3}\mathbf{r,} \label{br6a}%
\end{equation}
where $\mathbf{E},\mathbf{H}$ are the corresponding intensities of electric
and magnetic field. Due to the symmetry of current (\ref{br4}) that generates
these fields, the corresponding AM satisfies%
\begin{equation}
\mathbf{J}^{\gamma}=0, \label{BR5}%
\end{equation}
the proof is given in Appendix. This relation follows only from the angular
terms in the wave function (\ref{rs1}) and does not depend on the radial ones.
In other words, the relation (\ref{BR5}) holds not only for a free electron,
but also for an electron bound in the hydrogen atom. The result represents a
mean value, which is not influenced by the fluctuations generated by single
$\gamma$. So, this calculation suggests the integral AM of the cloud of
virtual $\gamma$ is zero despite the fact that AM of its source, the electron
in a state (\ref{rs1}), is not zero. While the free electron emits and
reabsorbs virtual photons by itself, the electron bounded in hydrogen atom in
addition exchanges (emits and absorbs) virtual photons with the proton. Since
the AM of the electromagnetic field generated by the proton is zero as well,
the total AM of hydrogen will be given only by AMs of the electron and proton,
without contribution of the electromagnetic field generated by both the particles.

Similar arguments can be relevant also for atoms in general and perhaps for
the nucleons bound in a nucleus. This would suggest the virtual particles
mediating the binding of nucleons also do not contribute to the resulting spin
of nucleus, which must be always integer or half-integer.

Our approach (A) has a common basis with the QED calculation (B) suggested in
Ref. \cite{Liu:2014fxa}, since both the approaches follow from the general QED
equations (1),(6) and (7) in the last reference. Despite that, there are some
differences between them, the most apparent are as follows.

(A) The approach is semiclassical only and the AMs are directly related to the
electron wavefunction $\psi$ and the classical electromagnetic field $A_{\mu}$
generated by the electric current $\bar{\psi}\gamma_{\mu}\psi$. The preferred
reference frame is the frame defined by the spinor spherical harmonics
(\ref{rs1}) or the rest frame of the defined composite system (e.g. atom).
This simplified treatment allows us to obtain the relation (\ref{BR5}) but
without a decomposition into the spin and OAM parts.

(B) The study is focused on the fundamental problem of the AM decomposition in
quantum field theory and explicit calculation is performed for the QED. The
preferred reference frame is the infinite momentum frame. The light-front
formalism is adopted to achieve a compatibility with the standard formulation
of the parton model. The result on the total boson (photon) AM seems be also
rather small, $S_{b}+L_{b}\simeq\mathcal{O}(e^{2}).$

So both the approaches, adopting rather different formalism and
correspondingly also some different assumptions, are not contradictory. From a
phenomenological point of view, they are complementary.

\subsection{Spin of quark}

\begin{figure}[t]
\includegraphics[width=9cm]{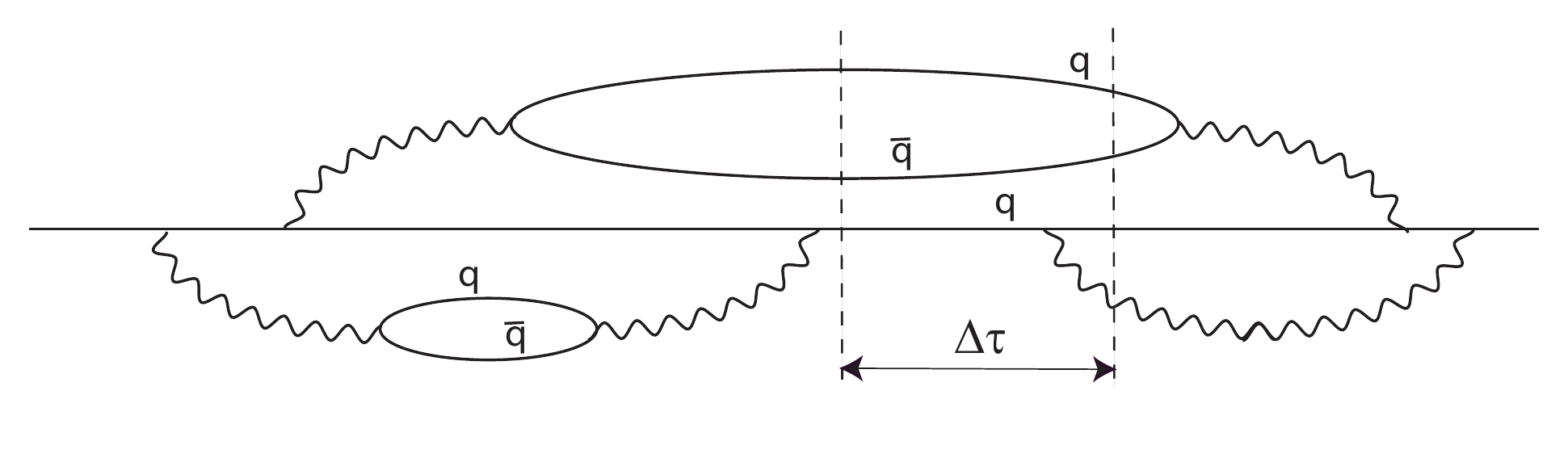}\caption{Scale dependent image of a
valence quark, see text.}%
\label{fgr2}%
\end{figure}

The situation with quarks inside a nucleon is more complicated. The quark at
different scales is sketched in Fig. \ref{fgr1} (in which the bare electron
surrounded by virtual cloud of $\gamma$ and $e^{-}e^{+}$ pairs is now replaced
by the bare quark with a cloud of virtual $g$ \ and $q\bar{q}$ $\ $\ pairs).
The terminology is as follows:

i) The bare quark can be identified with the current quark, which can be
described by the distribution functions $q^{a}(x)$ defined in the quark-parton
model. They are related to the sets of quarks and antiquarks in the figure for
$Q^{2}\rightarrow\infty.$

ii) The constituent quark can be identified with the dressed quark at a low
$Q^{2}$ scale.

iii) The valence quark can be identified with the set of quarks, from which
the cloud of virtual $q\bar{q}$ pairs and gluons is separated off. In the
figure the valence quarks are represented by the central spots. Strictly
speaking, depending on the scale, valence and sea quarks may not be clearly
distinguishable. In a short time interval $\Delta\tau$, a quark from the
virtual $q\bar{q}$ pair is indistinguishable from the source, valence quark,
see Fig. \ref{fgr2}. However, the usual definition in terms of the quark-parton model
distributions
\begin{equation}
q_{val}^{a}(x,Q^{2})=q^{a}(x,Q^{2})-\bar{q}^{a}(x,Q^{2}) \label{br7}%
\end{equation}
is unambiguous. For quarks the parameter $Q^{2}$ represents the
renormalization scale, but also the DIS parameter ($-Q^{2}=$ photon
four-momentum square) or equivalently a scale of the space-time domain inside
which the photon absorption takes place \cite{Zavada:2013ola}.

Now, we can put the question b) for the quarks bound inside the proton: How
much the field of virtual gluons generated by the valence quarks and the sea
of virtual $q\bar{q}$ pairs created by the gluons contribute to the proton
spin? This question is being studied in the experiments, which measure
contribution of the gluons and sea quarks to the proton spin. Available data
from the experiments COMPASS \cite{Adolph:2012vj} and HERMES
\cite{Airapetian:2010ac} suggest rather small gluon contribution, in fact the
data are consistent with zero within statistical errors.\ The very recent
results of the RHIC experiments
\ \cite{Adare:2014hsq,Adamczyk:2014ozi,deFlorian:2014yva,Nocera:2014gqa,Bourrely:2014uha}
suggest a positive gluon contribution which, however still cannot fully
compensate for a small quark contribution to the proton spin.

These results can be interpreted in the framework of covariant approach
presented in Ref. \cite{Zavada:2013ola}. In the paper we studied the
relativistic interplay between the quark spins and OAMs, which collectively
contribute to the proton spin. \ The simplest scenario assuming

1) the quarks are in the state $j=1/2,$ see Eq. 113\cite{Zavada:2013ola},

2) mass of quarks can be neglected, $\left\langle m/\epsilon\right\rangle
\rightarrow0,$

3) there is no gluon contribution, i.e. proton spin $J=1/2$ is generated only
by the AM of quarks, see Eq. 109\cite{Zavada:2013ola}

\noindent gave a prediction for the contribution of the quark spins in DIS
region,
\begin{equation}
\Delta\Sigma=\frac{1}{3}, \label{br2}%
\end{equation}
while the \textquotedblleft missing\textquotedblright\ part of the proton spin
is fully compensated by the quark OAM. This prediction fits the data
\cite{Alexakhin:2006vx,Airapetian:2007mh,Alekseev:2010ub} surprisingly well.

However, in a more general case, if only condition 1) is assumed, then AM of
each quark consists of the spin and OAM part%
\begin{equation}
\left\langle s_{z}\right\rangle =\frac{1+2\tilde{\mu}}{3}j_{z},\quad
\left\langle l_{z}\right\rangle =\frac{2-2\tilde{\mu}}{3}j_{z},\quad
\frac{\left\langle l_{z}\right\rangle }{\left\langle s_{z}\right\rangle
}=\frac{2-2\tilde{\mu}}{1+2\tilde{\mu}}, \label{br8z}%
\end{equation}
where $j_{z}=\pm1/2$, see Eqs. (17), (22) in Ref. [1]. It is very important
result, which easily follows from the algebra of spinor spherical harmonics
representing the solutions of Dirac equation. The ratio $\tilde{\mu
}=\left\langle m/\epsilon\right\rangle $ here plays a crucial role, since it
controls a "contraction" of the spin component, which is compensated by the
OAM. It is a quantum mechanical effect of relativistic kinematics.
Mathematically, a small $\tilde{\mu}$ means the lower component of Dirac
spinor is important. This is also the case in \cite{Ma:1992sj,Qing:1998at},
where with some distinction of formalism and paradigm the authors come to
similar results on the spin of quark bound in the proton. The effect is
strongly correlated with the transversal motion of quarks inside the nucleon
\cite{Zhang:2012sta}.

The relations (\ref{br8z}) imply that the total AM of a composite system of
quarks with $j_{1}=j_{2}=j_{3}=...=1/2$ reads
\begin{equation}
J^{q}=\left\langle \mathbb{S}_{z}\right\rangle +\left\langle \mathbb{L}%
_{z}\right\rangle , \label{br8}%
\end{equation}
where the ratio of the total spin $\left\langle \mathbb{S}_{z}\right\rangle $
and OAM $\left\langle \mathbb{L}_{z}\right\rangle $ is the same as for the
one-quark states above:
\begin{equation}
\frac{\left\langle \mathbb{L}_{z}\right\rangle }{\left\langle \mathbb{S}%
_{z}\right\rangle }=\frac{2-2\tilde{\mu}}{1+2\tilde{\mu}}. \label{br9}%
\end{equation}
Further, if the proton spin consists of the quark and gluon contributions, one
can write%
\begin{equation}
\frac{1}{2}=J^{q}+J^{g};\qquad J^{q}=\frac{1}{2}\varkappa,\qquad J^{g}%
=\frac{1}{2}\left(  1-\varkappa\right)  . \label{br10}%
\end{equation}
With the use of Eqs.(\ref{br8})-(\ref{br10}) one gets
\begin{align}
\left\langle \mathbb{S}_{z}\right\rangle +\left\langle \mathbb{L}%
_{z}\right\rangle  &  =\left\langle \mathbb{S}_{z}\right\rangle \left(
1+\frac{2-2\tilde{\mu}}{1+2\tilde{\mu}}\right)  =\frac{1}{2}\varkappa
,\label{br11}\\
\varkappa &  =1-2J^{g},\nonumber
\end{align}
which after replacing $\left\langle \mathbb{S}_{z}\right\rangle =\Delta
\Sigma/2$ gives%
\begin{equation}
\Delta\Sigma=\frac{1}{3}\left(  1-2J^{g}\right)  \left(  1+2\tilde{\mu
}\right)  . \label{br14}%
\end{equation}
In a higher approximation, if we admit an admixture of the quark states with
$j\geq3/2,$ the relation is modified as%
\begin{equation}
\Delta\Sigma\lesssim\frac{1}{3}\left(  1-2J^{g}\right)  \left(  1+2\tilde{\mu
}\right)  . \label{br14a}%
\end{equation}

\section{Discussion and conclusion}

\label{proton}
\begin{figure}[t]
\includegraphics[width=8cm]{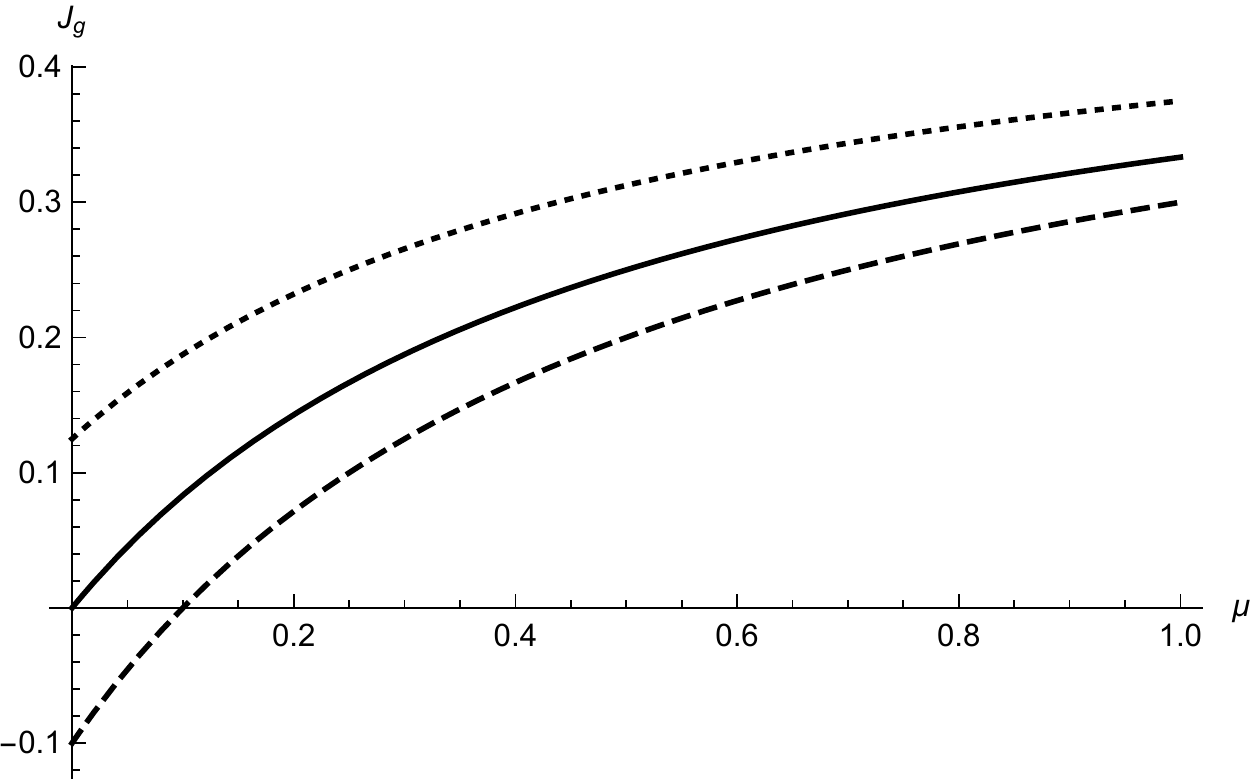}\caption{Dependence of\ $\Delta\Sigma$ on
$J_{g}$ and $\tilde{\mu}$. The dotted, full and dashed lines correspond to
$\Delta\Sigma=0.25,0.33$ and $0.4$ respectively.}%
\label{fgr3}%
\end{figure}

The relation (\ref{br14}) means the quark spin content \ depends
on two parameters, the gluon contribution $J^{g}$\ and the quark effective
mass ratio $\tilde{\mu}$, which affects proportion of the quark OAM. It
follows from the algebra of spinor spherical harmonics and from general rules
of AM composition in the system of quarks and gluons with the total spin
$J=1/2$. The dependence is demonstrated in Fig. \ref{fgr3}. One can observe:

a) $\Delta\Sigma\leq1/3$ corresponds to $J^{g}\geq0$ for any $1\geq\tilde{\mu
}\geq0$. A special case $\Delta\Sigma=1/3$ and $\tilde{\mu}\rightarrow0$
implies $J^{g}\rightarrow0$.

b) $\Delta\Sigma>1/3,$ then the sign of $J^{g}$ depends on $\tilde{\mu}.$
Apparently, $J^{g}<0$ would imply $J^{q}>1/2.$

In this way the COMPASS and HERMES data
\cite{Alexakhin:2006vx,Airapetian:2007mh,Alekseev:2010ub} giving $\Delta
\Sigma\approx1/3,$ can be compatible also with a positive gluon contribution
$J^{g}$ suggested by the recent data on RHIC. Note that positive $J^{g}%
$\ correlates with a positive quark effective mass ratio $\tilde{\mu}$. Can we
somehow estimate this parameter? In the proton rest frame a quark has momentum
$\left\langle k\right\rangle =\sqrt{3/2}\left\langle k_{T}\right\rangle $ and
there are independent ways to estimate it, for example:

1) If we take proton diameter \ $d_{p}=0.84$fm, then the uncertainty relation
gives for the corresponding momentum roughly $k\approx230$MeV.

2) The analysis \cite{Chay:1991nh, Adams:1993hs,
Anselmino:2006yc,Schweitzer:2010tt} of the data on the azimuthal asymmetry in
semi-inclusive DIS suggest $\left\langle k_{T}\right\rangle \approx400-600$MeV.

3) The statistical approaches
\cite{Cleymans:1986gy,Bhalerao:1996fc,Bourrely:2001du} suggests $\left\langle
k\right\rangle \approx40-100$MeV and a similar value was obtained in Ref.
\cite{Zavada:2007ww} for valence quarks, see also discussion in Ref.
\cite{Efremov:2010mt,Zavada:2011cv}.

Therefore, if one suppose the quark effective mass of the order MeV, then the
parameter $\tilde{\mu}\lesssim0.1$, which gives a similar upper limit on the
gluon contribution, $J^{g}\lesssim0.1$.\ Finally, the spin contribution of the
sea quarks is known to be small or compatible with zero \cite{Alekseev:2010ub}%
. This can confirm the expectation that the sea quark contribution correlates
with the gluon contribution.

In the present approach the quark effective mass ratio $\tilde{\mu}$ and the
gluon AM contribution $J^{g}$ are free, phenomenological parameters
constrained by the relation (\ref{br14}). At the same time corresponding
scale-dependent parameters, for example the quark effective masses, are (at
least in principle) calculable in QCD \cite{IOFFE}. However, due to
nonperturbative aspect of related task, the real calculation can be extremely
difficult. That is why our approach based on the covariant quark-parton model
can be a useful supplement to the exact but more complicated theory of the
nucleon spin structure based on pure QCD.

To conclude, interpretation of the available sets of experimental data in
framework of the covariant approach suggests an important role of the quark
OAM for the creation of the proton spin on the scale $Q^{2}$ defined by the
data. The data of experiments COMPASS and HERMES on the quark and gluon
contribution to the proton spin are fully compatible with our approach. At the
same time, a limited positive gluon contribution does not contradict the
covariant approach. However, the precise data on $J^{g}$ are still missing, so
the existing experimental data do not disprove the hypothesis $J^{g}\approx0$
based on the analogy with AM of virtual photons given by Eq. (\ref{BR5}).

\begin{acknowledgments}
This work was supported by the project LG130131 of the MEYS (Czech Republic).
I am grateful to Anatoli Efremov, Peter Filip, Bo-Qiang Ma and Oleg Teryaev
for useful discussions and valuable comments.
\end{acknowledgments}

\appendix

\section{Proof of the relation (\ref{BR5})}

The current (\ref{br4}) generates the electric and magnetic field%
\begin{align}
\mathbf{E}(\mathbf{r})  &  =\int I_{0}(\mathbf{r}^{\prime})\frac
{\mathbf{r}-\mathbf{r}^{\prime}}{\left\vert \mathbf{r}-\mathbf{r}^{\prime
}\right\vert ^{3/2}}d^{3}\mathbf{r}^{\prime},\label{a1}\\
\mathbf{H}(\mathbf{r})  &  =\int\mathbf{I(\mathbf{r}^{\prime})\times}%
\frac{\mathbf{r}-\mathbf{r}^{\prime}}{\left\vert \mathbf{r}-\mathbf{r}%
^{\prime}\right\vert ^{3/2}}d^{3}\mathbf{r}^{\prime}. \label{a2}%
\end{align}
If we define%
\begin{align}
\mathbf{W}^{X}(\mathbf{r})  &  =\int\frac{h_{X}(r\mathbf{^{\prime})}%
\rho_{j,j_{z}}\left(  \cos\theta^{\prime}\right)  \mathbf{r}^{\prime}%
}{\left\vert \mathbf{r}-\mathbf{r}^{\prime}\right\vert ^{3/2}}d^{3}%
\mathbf{r}^{\prime},\label{a3}\\
S(\mathbf{r})  &  =\int\frac{h_{I}(r\mathbf{^{\prime})}\rho_{j,j_{z}}\left(
\cos\theta^{\prime}\right)  }{\left\vert \mathbf{r}-\mathbf{r}^{\prime
}\right\vert ^{3/2}}d^{3}\mathbf{r}^{\prime},\nonumber
\end{align}
where $X=I,II$, then with the use of (\ref{br4}),(\ref{br6}) we get:%
\begin{equation}
\mathbf{E}(\mathbf{r})=-\mathbf{W}^{I}(\mathbf{r})+S(\mathbf{r})\mathbf{r}%
,\qquad\mathbf{H}(\mathbf{r})=\mathbf{W}^{II}(\mathbf{r})\times\mathbf{r}.
\label{a5}%
\end{equation}
In terms of spherical coordinates%
\begin{equation}
r_{1}=r\sin\theta\cos\varphi,\ r_{2}=r\sin\theta\sin\varphi,\ r_{3}%
=r\cos\theta\label{a6}%
\end{equation}
we have%
\begin{align}
S(\mathbf{r})  &  =\int\frac{h_{I}(r\mathbf{^{\prime})}\rho_{j,j_{z}}\left(
\cos\theta^{\prime}\right)  }{\left(  r^{2}+r^{\prime^{2}}-2rr^{\prime}\left(
\sin\theta\sin\theta^{\prime}\cos\left(  \varphi-\varphi^{\prime}\right)
+\cos\theta\cos\theta^{\prime}\right)  \right)  ^{3/2}}r^{\prime^{2}}%
\sin\theta^{\prime}d\varphi^{\prime}d\theta^{\prime}dr^{\prime},\label{a7}\\
\mathbf{W}^{X}(\mathbf{r})  &  =\int\frac{h_{X}(r\mathbf{^{\prime})}%
\rho_{j,j_{z}}\left(  \cos\theta^{\prime}\right)  \mathbf{r}^{\prime}}{\left(
r^{2}+r^{\prime^{2}}-2rr^{\prime}\left(  \sin\theta\sin\theta^{\prime}%
\cos\left(  \varphi-\varphi^{\prime}\right)  +\cos\theta\cos\theta^{\prime
}\right)  \right)  ^{3/2}}r^{\prime^{2}}\sin\theta^{\prime}d\varphi^{\prime
}d\theta^{\prime}dr^{\prime}. \label{a8}%
\end{align}
Obviously $\mathbf{S}(\mathbf{r})$\ does not depend on $\varphi$ so we have \
\begin{equation}
S(\mathbf{r})=S(r,\theta). \label{a9}%
\end{equation}
In the second integral, after substitution $\psi=\varphi^{\prime}-\varphi$ we
replace correspondingly in $\mathbf{r}^{\prime}:$
\begin{align}
x^{\prime}  &  =r^{\prime}\sin\theta^{\prime}\cos\varphi^{\prime}\rightarrow
r^{\prime}\sin\theta^{\prime}\left(  \cos\psi\cos\varphi-\sin\psi\sin
\varphi\right)  ,\label{a10}\\
y^{\prime}  &  =r^{\prime}\sin\theta^{\prime}\sin\varphi^{\prime}\rightarrow
r^{\prime}\sin\theta^{\prime}\left(  \cos\psi\sin\varphi+\sin\psi\cos
\varphi\right)  \label{a11}%
\end{align}
and instead of (\ref{a8}) we obtain%
\begin{align}
W_{1}^{X}(\mathbf{r})  &  =\int\frac{h_{X}(r\mathbf{^{\prime})}\rho_{j,j_{z}%
}\left(  \cos\theta^{\prime}\right)  r^{\prime}\sin\theta^{\prime}\left(
\cos\psi\cos\varphi-\sin\psi\sin\varphi\right)  }{\left(  r^{2}+r^{\prime^{2}%
}-2rr^{\prime}\left(  \sin\theta\sin\theta^{\prime}\cos\psi+\cos\theta
\cos\theta^{\prime}\right)  \right)  ^{3/2}}r^{\prime^{2}}\sin\theta^{\prime
}d\psi d\theta^{\prime}dr^{\prime},\label{a12}\\
W_{2}^{X}(\mathbf{r})  &  =\int\frac{h_{X}(r\mathbf{^{\prime})}\rho_{j,j_{z}%
}\left(  \cos\theta^{\prime}\right)  r^{\prime}\sin\theta^{\prime}\left(
\cos\psi\sin\varphi+\sin\psi\cos\varphi\right)  }{\left(  r^{2}+r^{\prime^{2}%
}-2rr^{\prime}\left(  \sin\theta\sin\theta^{\prime}\cos\psi+\cos\theta
\cos\theta^{\prime}\right)  \right)  ^{3/2}}r^{\prime^{2}}\sin\theta^{\prime
}d\psi d\theta^{\prime}dr^{\prime},\label{a13}\\
W_{3}^{X}(\mathbf{r})  &  =\int\frac{h_{X}(r\mathbf{^{\prime})}\rho_{j,j_{z}%
}\left(  \cos\theta^{\prime}\right)  r^{\prime}\cos\theta^{\prime}}{\left(
r^{2}+r^{\prime^{2}}-2rr^{\prime}\left(  \sin\theta\sin\theta^{\prime}\cos
\psi+\cos\theta\cos\theta^{\prime}\right)  \right)  ^{3/2}}r^{\prime^{2}}%
\sin\theta^{\prime}d\psi d\theta^{\prime}dr^{\prime}. \label{a14}%
\end{align}
Since in general
\begin{equation}
\int_{-\pi}^{\pi}f_{even}(\psi)\sin\psi d\psi=0, \label{a15}%
\end{equation}
where $f_{even}(\psi)=f_{even}(-\psi),$ then the second term in (\ref{a12}),(
\ref{a13}) vanishes and the expressions are simplified as%
\begin{equation}
W_{1}^{X}(\mathbf{r})=W^{X}(r,\theta)r_{1},\qquad W_{2}^{X}(\mathbf{r}%
)=W^{X}(r,\theta)r_{2},\qquad W_{3}^{X}(\mathbf{r})=W_{3}^{X}(r,\theta),
\label{a16}%
\end{equation}
where%
\begin{align}
W^{X}(r,\theta)  &  =\frac{1}{r\sin\theta}\int\frac{h_{X}(r\mathbf{^{\prime}%
)}\rho_{j,j_{z}}\left(  \cos\theta^{\prime}\right)  r^{\prime}\sin
\theta^{\prime}\cos\psi}{\left(  r^{2}+r^{\prime^{2}}-2rr^{\prime}\left(
\sin\theta\sin\theta^{\prime}\cos\psi+\cos\theta\cos\theta^{\prime}\right)
\right)  ^{3/2}}r^{\prime^{2}}\sin\theta^{\prime}d\psi d\theta^{\prime
}dr^{\prime},\label{a17}\\
W_{3}^{X}(r,\theta)  &  =\int\frac{h_{X}(r\mathbf{^{\prime})}\rho_{j,j_{z}%
}\left(  \cos\theta^{\prime}\right)  r^{\prime}\cos\theta^{\prime}}{\left(
r^{2}+r^{\prime^{2}}-2rr^{\prime}\left(  \sin\theta\sin\theta^{\prime}\cos
\psi+\cos\theta\cos\theta^{\prime}\right)  \right)  ^{3/2}}r^{\prime^{2}}%
\sin\theta^{\prime}d\psi d\theta^{\prime}dr^{\prime}. \label{a18}%
\end{align}

After inserting from (\ref{a5}) into (\ref{br6a}) we integrate the AM density%
\begin{equation}
\mathbf{j}^{\gamma}=\mathbf{r}\times\left(  \left(  -\mathbf{W}^{I}%
(\mathbf{r})+S(\mathbf{r})\mathbf{r}\right)  \times\left(  \mathbf{W}%
^{II}(\mathbf{r})\times\mathbf{r}\right)  \right)  , \label{a19}%
\end{equation}
which with the use of (\ref{a9}), (\ref{a16}) gives%
\begin{align}
\mathbf{j}^{\gamma}  &  =\left\{
\begin{array}
[c]{c}%
\left(  W^{I}W^{II}\left(  r_{1}^{2}r_{2}r_{3}+r_{2}^{3}r_{3}\right)
+W^{II}W_{3}^{I}r_{2}r_{3}^{2}-W^{I}W_{3}^{II}\left(  r_{1}^{2}r_{2}+r_{2}%
^{3}\right)  -W_{3}^{I}W_{3}^{II}r_{2}r_{3}\right)  ,\\
\left(  -W^{I}W^{II}\left(  r_{1}r_{2}^{2}r_{3}+r_{1}^{3}r_{3}\right)
-W^{II}W_{3}^{I}r_{1}r_{3}^{2}+W^{I}W_{3}^{II}\left(  r_{1}r_{2}^{2}+r_{1}%
^{3}\right)  +W_{3}^{I}W_{3}^{II}r_{1}r_{3}\right)  ,\\
0
\end{array}
\right\} \label{a20}\\
&  +Sr^{2}\left\{
\begin{array}
[c]{c}%
\left(  -W^{II}r_{2}r_{3}+W_{3}^{II}r_{2}\right)  ,\\
\left(  W^{II}r_{1}r_{3}-W_{3}^{II}r_{1}\right)  ,\\
0
\end{array}
\right\}  .\nonumber
\end{align}
These terms depend on $\varphi$\ only via coordinates $r_{1}$ and $r_{2}$
\ (\ref{a6}). Since each term involves just one odd power of $r_{1}\sim
\cos\varphi$ or $r_{2}\sim\sin\varphi$, the corresponding integral satisfies
(\ref{BR5}).


\begin{thebibliography}{99}                                                                                               %


\bibitem {Zavada:2013ola}P.~Zavada,
Phys.\ Rev.\ D \textbf{89}, 014012 (2014) .



\bibitem {Adare:2014hsq}A.~Adare \textit{et al.} [PHENIX Collaboration],
Phys.\ Rev.\ D \textbf{90}, no. 1, 012007 (2014) .



\bibitem {Adamczyk:2014ozi}L.~Adamczyk \textit{et al.} [STAR Collaboration],
Phys.\ Rev.\ Lett.\ \textbf{115}, no. 9, 092002 (2015) .



\bibitem {deFlorian:2014yva}D.~de Florian, R.~Sassot, M.~Stratmann and
W.~Vogelsang,
Phys.\ Rev.\ Lett.\ \textbf{113}, 012001 (2014) .



\bibitem {Nocera:2014gqa}E.~R.~Nocera \textit{et al.} [NNPDF Collaboration],
Nucl.\ Phys.\ B \textbf{887}, 276 (2014) .



\bibitem {Bourrely:2014uha}C.~Bourrely and J.~Soffer,
Phys.\ Lett.\ B \textbf{740}, 168 (2015) .



\bibitem {Liu:2014fxa}T.~Liu and B.~-Q.~Ma,
Phys.\ Rev.\ D \textbf{91}, 017501 (2015) .



\bibitem {Brodsky:2000ii}S.~J.~Brodsky, D.~S.~Hwang, B.~-Q.~Ma and
I.~Schmidt,
Nucl.\ Phys.\ B \textbf{593}, 311 (2001) .

\bibitem {Burkardt}M. Burkardt and Hikmat BC, Phys.\ Rev.\ D \textbf{79},
071501(R) (2009).



\bibitem {Adolph:2012vj}C.~Adolph \textit{et al.} [COMPASS Collaboration],
Phys.\ Lett.\ B \textbf{718}, 922 (2013) .



\bibitem {Airapetian:2010ac}A.~Airapetian \textit{et al.} [HERMES
Collaboration],
JHEP \textbf{1008}, 130 (2010) .



\bibitem {Alexakhin:2006vx}V.~Y.~.Alexakhin \textit{et al.} [COMPASS
Collaboration],
Phys.\ Lett.\ B \textbf{647}, 8 (2007) .



\bibitem {Airapetian:2007mh}A.~Airapetian \textit{et al.} [HERMES
Collaboration],
Phys.\ Rev.\ D \textbf{75}, 012007 (2007) .



\bibitem {Alekseev:2010ub}M.~G.~Alekseev \textit{et al.} [COMPASS
Collaboration],
Phys.\ Lett.\ B \textbf{693}, 227 (2010) .



\bibitem {Ma:1992sj}B.~-Q.~Ma and Q. -R Zhang,
Z.\ Phys.\ C \textbf{58}, 479 (1993) .



\bibitem {Qing:1998at}D.~Qing, X.~-S.~Chen and F.~Wang,
Phys.\ Rev.\ D \textbf{58}, 114032 (1998) .



\bibitem {Zhang:2012sta}X.~Zhang and B.~-Q.~Ma,
Phys.\ Rev.\ D \textbf{85}, 114048 (2012) .



\bibitem {Anselmino:2006yc}M.~Anselmino, A.~Efremov, A.~Kotzinian and
B.~Parsamyan,
Phys.\ Rev.\ D \textbf{74}, 074015 (2006) .

\bibitem {Chay:1991nh}J.~Chay, S.~D.~Ellis and W.~J.~Stirling,
Phys.\ Rev.\ D \textbf{45}, 46 (1992).


\bibitem {Adams:1993hs}M.~R.~Adams \textit{et al.} [E665 Collaboration],
Phys.\ Rev.\ D \textbf{48}, 5057 (1993).

\bibitem {Schweitzer:2010tt}P.~Schweitzer, T.~Teckentrup and A.~Metz,
Phys.\ Rev.\ D \textbf{81}, 094019 (2010) .



\bibitem {Cleymans:1986gy}J.~Cleymans and R.~L.~Thews,
Z.\ Phys.\ C \textbf{37}, 315 (1988).



\bibitem {Bhalerao:1996fc}R.~S.~Bhalerao,
Phys.\ Lett.\ B \textbf{380}, 1 (1996) [Phys.\ Lett.\ B \textbf{387}, 881
(1996)] .

\bibitem {Bourrely:2001du}C.~Bourrely, J.~Soffer and F.~Buccella,
Eur.\ Phys.\ J.\ C \textbf{23}, 487 (2002);
Mod.\ Phys.\ Lett.\ A \textbf{18}, 771 (2003);
Eur.\ Phys.\ J.\ C \textbf{41}, 327 (2005);
Mod.\ Phys.\ Lett.\ A \textbf{21}, 143 (2006);
Phys.\ Lett.\ B \textbf{648}, 39 (2007).



\bibitem {Zavada:2007ww}P.~Zavada,
Eur.\ Phys.\ J.\ C \textbf{52}, 121 (2007) .



\bibitem {Efremov:2010mt}A.~V.~Efremov, P.~Schweitzer, O.~V.~Teryaev and
P.~Zavada,
Phys.\ Rev.\ D \textbf{83}, 054025 (2011) .



\bibitem {Zavada:2011cv}P.~Zavada,
Phys.\ Rev.\ D \textbf{85}, 037501 (2012) .

\bibitem {IOFFE}B.L. Ioffe, V.S. Fadin and L.N. Lipatov, \textit{Quantum
chromodynamics: perturbative and nonperturbative aspects}, Cambridge
University Press, 2010, ISBN 978-0-521-63148-8.
\end{thebibliography}
\end{document}